\documentclass{aa}
\usepackage{graphicx}
\usepackage{txfonts}
\begin{document}

   \title{Investigating the vertical distribution of the disk as a function of radial action}


   \author{Yunpeng Jia
          \inst{1}\thanks{jiayunpeng11@mails.ucas.ac.cn}
          \and
          Yuqin Chen\inst{2,3}
          \and
          Cuihua Du\inst{3}
          \and
          Gang Zhao\inst{2,3}
          }

   \institute{School of Physics, Shangqiu Normal University, Shangqiu, 476000, P. R. China
         \and
            CAS Key Laboratory of Optical Astronomy, National Astronomical Observatories, 
Chinese Academy of Sciences, Beijing, 100101, P. R. China
\and 
School of Astronomy and Space Science, University of Chinese Academy of Sciences, Beijing 100049, P. R. China
}


 
  \abstract
  {}
   {As heating processes can broaden the distributions of radial actions and the vertical distributions of the Galactic disks, we investigate the vertical distribution of the Galactic disks as a function of radial action based on Apache Point Observatory Galactic Evolution Experiment(APOGEE) and Gaia data in order to deepen our understanding of the formation and heating history of the Galactic disks.}
   {The vertical distributions of the thin and thick Galactic disks defined in the chemical plane were fitted with a simple exponential function with a free parameter of scale height in different radial action ranges. Therefore, we were able to analyze the scale height as a function of radial action for different disk populations.
   }
   {We find that the distributions of radial action for both the thin and thick disks can be approximately described by pseudo-isothermal distributions, which give a statistical measurement for the temperature of the disk as indicated by the mean radial action of the star sample. Estimations of the scale heights in different radial action ranges for these pseudo-isothermal distributions of the disks seem to show fixed relationships between radial action $J_R$ and scale height $h$. We describe these relationships with a two-parameter function of $h=\sqrt{J_R/a}+b$, where $a$ and $b$ are free parameters. When testing with a three-parameter function of $h=\sqrt[\alpha]{J_R/a}+b$, we find that this two-parameter function describes the thin disk well, but we note the function should be used with care for the thick disk. When comparing the best-fit relationships between the inner and outer disk for both of the thin and thick disks, we find that the relationships are nearly the same for the thin disks but are different for the thick disks. The inner thick disk shows a nearly flattened relationship, while the outer thick disk presents a gradually increasing relationship. This work highlights an alternative way to unveil the heating history of the disks by investigating the relationship between scale height and radial action, as these relationships encode the formation and heating history of the Galactic disks.
   }  
  {}

   \keywords{Galaxy: disk  --  Galaxy: formation --  Galaxy: structure  --  Galaxy: fundamental parameters}

   \titlerunning{Investigation on the scale height as a function of radial action}
   \maketitle
   
%

\section{Introduction}

It has long been recognized that Galactic disk stars could be scattered by giant molecular clouds \citep[][]{Spitzer1953,Lacey1984,Hanninen2002} or
transient spiral structures \citep{Barbanis1967,Lynden-Bell1972,Carlberg1985}, or 
by both \citep[][or see \citealt{Sellwood2014}]{Julian1966}, 
increasing the random motion (or radial action) of the stars.
Such processes could drive secular changes in the energy and angular momentum (or guiding radius) of disk stars and result in a gradual increase in velocity dispersion with age for subgroups of disk stars. For this reason, such processes are loosely referred to as heating processes.

These kinds of dynamical processes are also found to be responsible for the formation of  the Galactic thick disk because they
are able to bring disk stars from a small vertical extent into a large one, allowing the thin disk stars to develop into the thick disk stars.
For example, the heating of a preexisting thin disk through minor mergers \citep[e.g.,][]{Quinn1993, Villalobos2008} is a mechanism widely used to explain the origin of the Galactic thick disk. However, there are other mechanisms that allow the Galactic thick disk to form:  
i) accretion from disrupted satellites \citep{Abadi2003} and 
ii) in situ triggered star formation with various causes \citep[e.g.,][]{Brook2004, Brook2005, Bournaud2009, Bird2013, Bird2021},
which form the disk in an "upside-down" fashion (i.e., old stars form in a relatively thick layer, while young stars form in a thin layer).
In addition, as evidenced by previous works \citep[e.g.,][]{Nissen-Schuster2010,Nissen-Schuster2011,Belokurov2018,Haywood2018,Di_Matteo2019}, major merger events are another possible mechanism  contributing to the formation of the thick disk stars, which is supported by the work of \citet{Helmi18,Gallart2019,Zhao21}.

Previous research has shown that spiral structures do not always heat the Galactic disks. The pioneering work of \citet{Sellwood2002} demonstrated that stars
near corotation of a spiral pattern can vary their angular momenta without heating, which is a process known as radial migration.
 Further investigations have shown that radial migration can indeed occur due to a variety of sources, for example, spirals \citep[e.g.,][]{Roskar2012}, bars \citep[e.g.,][]{Friedli1994}, and satellites \citep[e.g.,][]{Quillen2009,Bird2012}, and conserves the vertical actions of stars \citep{Solway2012,Minchev2012a}. 
It should be noted that  kinematic heating can also occur during episodes of radial migration when resonant overlap is possible \citep{Minchev-Quillen2006,Minchev2010a,Minchev2011,Daniel2019}.
 Radial migration has been found to be able to account for the formation of the
 Galactic thick disk by the pioneering work of \citet{schonrich-binney2009} and further supported by subsequent research \citep[e.g.,][]{Schonrich2009a,Loebman2011,Roskar2013,Sharma2021}, though these works are  controversial because radial migration is found to have little effect on the thickening of the disk \citep{Minchev2012a, Halle2015}. However, \citet{Schonrich2017} stated that the outward radial migration process under action conservation can achieve larger scale heights, which makes this mechanism a still possible scenario for the origin of the Galactic thick disk.
Moreover, kinematic heating caused by radial migration \citep{Daniel2019} increases radial action, which could affect the
vertical distribution of disk stars and partly contribute to the formation of the Galactic thick disk.

Although there are different mechanisms related to the origin of the Galactic thick disk, it is helpful to investigate whether the thick disk stars are born thick or thin in order to partly distinguish the mechanism that dominates the formation of the Galactic thick disk. The cosmological simulations that have been run with different galaxy formation codes fail to give a consistent result for this question but nonetheless highlight the role of the physical processes in star formation \citep{House2011}, for example, the recipe for feedback \citep{Miranda2016}. An observational work using more than a hundred star-forming galaxies to investigate the variation in ionized gas velocity dispersions with redshift indicates that the thick disk is born hot \citep{Wisnioski2015}, while the work of \citet{Fraternali2021} studying high-redshift galaxies with emission lines of cold interstellar matter reveals the contrary.

In the Milky Way, the lack of direct observational evidence  causes this topic to remain unresolved.
The main reason is that the estimation of birth sites for the disk stars is nontrivial, especially for the thick disk stars, and the radial migration process makes this situation even worse \citep[e.g.,][]{Wielen1996,Frankel2018,Minchev2018, Quillen2018, Chen2019, Feltzing2020}. 
However, measuring the thickness of a population of related stars with a variable that can characterize heating is informative to the study of the heating origin of the thick disk.
For this purpose, radial action is a suitable parameter, since it gradually increases by heating processes. Previous works have provided several ways to characterize these heating processes by using actions \citep[e.g.,][]{Binney2010,Binney2011,Ting2019,Frankel2020}, but there is  no direct measurement for the relation of radial action with the thickness of the Galactic disk. 

In this work, we investigate the relationship between radial action and the vertical distribution of the Galactic disk using data from the Apache Point Observatory Galactic Evolution Experiment (APOGEE) and Gaia with the purpose of deepening our understanding of the formation and heating history of the Galactic disks.
This paper is structured as follows: in Section 2, we briefly describe the data adopted in this work; the methods and results are presented in Section 3. Finally, a discussion and conclusions are given in Section 4.

\section{Data}
In this work, we used a sample of the Galactic disk stars similar to \citep[][hereafter J18]{Jia2018} that was obtained through
cross-matching between the 14th data release of APOGEE (APOGEE DR14) and  the second data release of Gaia (Gaia DR2). 
However, where the stellar heliocentric distances in J18  were computed by inverting the parallaxes
from Gaia, in this work we updated the distances and kinematical and orbital parameters using a Bayesian approach that 
follows \citet{Yan2019}. The details of parameter inference can be found in the above
literature and references therein. In this section, we summarize the key points.

The posterior probability distribution of the parameters (heliocentric distance, tangential speed, and travel direction)
were derived by multiplying a Gaussian distribution likelihood by a separable prior of parameters. 
The heliocentric distance prior is
an exponentially decreasing space density distribution \citep{Bailer-Jones2015} with a Galactic longitude- and latitude-dependent
length scale taken from \citet{Bailer-Jones2018}. The tangential speed prior is a beta distribution where
$\alpha=2$, $\beta=3,$ and $v_{max}=750$ $\rm km\, s^{-1}$. The travel direction prior is a uniform distribution.
We obtained the posterior probability distribution of radial velocity by
multiplying a Gaussian distribution likelihood by a uniform prior in order to further 
estimate the kinematical and orbital parameters.
The mean and standard deviation in Gaussian distribution of radial velocity were provided by
the values of radial velocity and their error in the APOGEE catalog.

We used EMCEE \citep{emcee} to draw random samples from the above posterior distributions for the parameters of heliocentric distance, 
tangential speed, travel direction, and radial velocity. The medians of the posterior samples were then used to derive the distance in Galactic cylindrical coordinates $R$ and $Z$
and the spatial velocities ($U,V,W$) in right-hand Cartesian coordinates. 
Coordinate transformation was adopted to obtain the cylindrical coordinate velocities
$V_R$ and $V_{\phi}$ after correcting the solar peculiar motion 
of $(U, V, W)=(11.1, 12.24, 7.25)\, \rm km\, s^{-1}$ \citep{Schonrich-Binney2010}.
Then, the guiding radius ($R_g$) and orbital parameters, such as peri-center ($R_p$), apo-center ($R_a$), and
eccentricity ($e=(R_a-R_p)/(R_a+R_p)$), were estimated with 
 \emph{galpy} \citep{Bovy2015} under the potential of  \emph{McMillan17} \citep{McMillan2017}. The actions were calculated with the Staeckel approximation \citep{Binney2012} under this potential.
In the calculation, we adopted the Gaia parallax zero point of -0.0523 mas \citep{Leung2019},
a distance from the Sun to the Galactic center of 8.21 kpc, and
a local standard of rest velocity of 233.1 $\rm km\, s^{-1}$ \citep{McMillan2017}.

The Galactic disk star sample was selected by using the same criteria as in J18, which 
excluded stars with metallicities less than $-1.0$ dex;   
 total velocities ($\sqrt{U^2+V^2+W^2}$ have been corrected for the solar peculiar motion) of less than 150 $\rm km\, s^{-1}$, to diminish the contamination from the halo; and stars at the 
fields of ($l$, $|b|$) $<$ (10$^\circ$, 10$^\circ$), to remove the bulge stars. 
At this stage, the star sample  comprised  149\,685 giant stars. However, there were
736 stars among them whose actions failed to be calculated under the potential of \emph{McMillan17}. Considering that these stars constituted less than one percent of our total star sample, we dropped them from the sample. The final number of stars in the sample adopted by this work is 148\,949.
        
\begin{figure}
\includegraphics[width=\hsize]{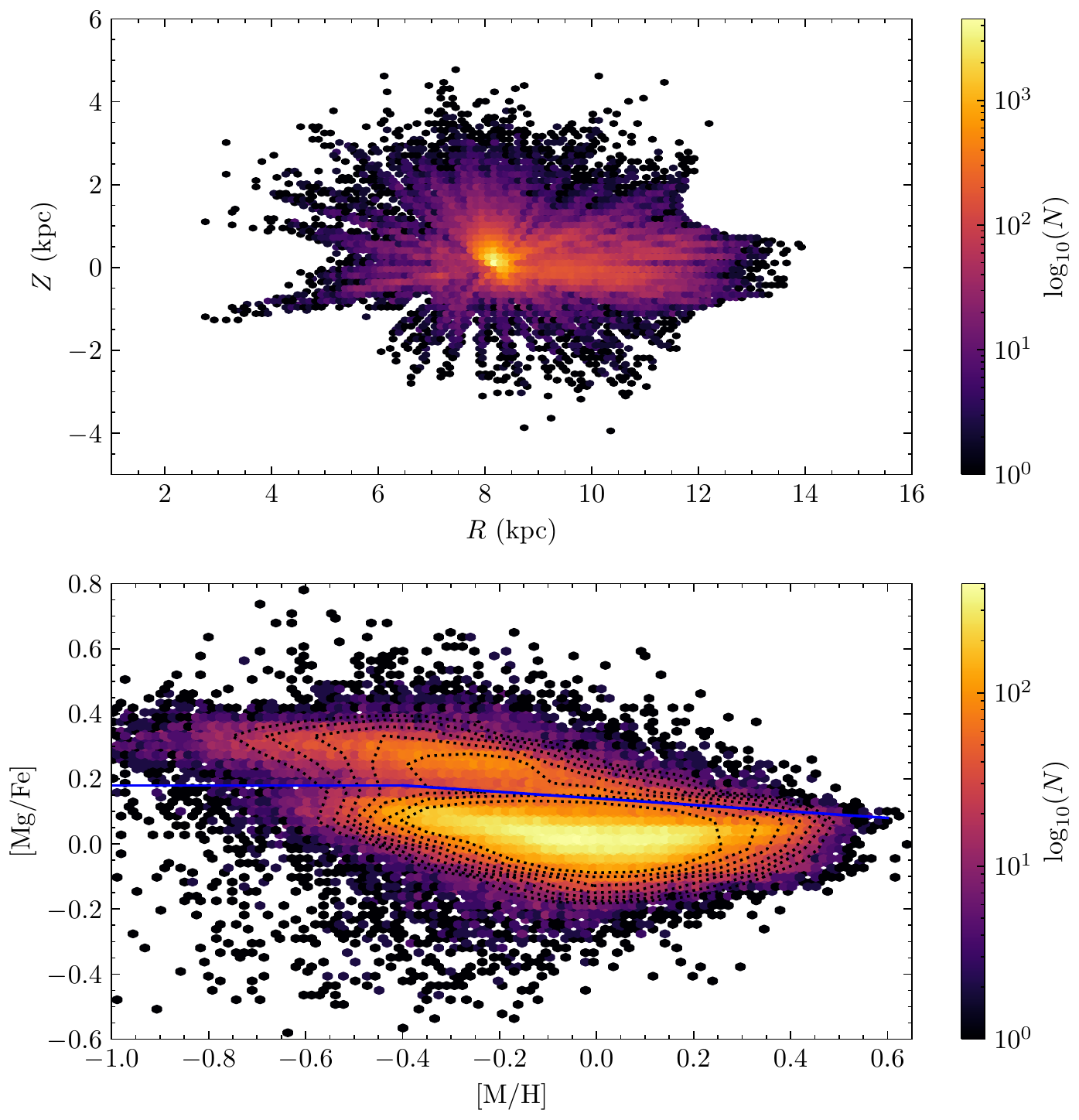}
\caption{ Upper panel: Distribution of the disk star sample in the $R$ versus $Z$ plane.
Bottom panel: Distribution of the star sample in the [M/H] versus [Mg/Fe] chemical plane. 
The solid line divides the thin disk from the thick disk.}\label{mh-mg}
\end{figure}

Most of the stars in this sample are located in a wide spatial range of the Galactic disk: $2<R<14$ kpc and $|Z|<5$ kpc (as shown in Fig.~\ref{mh-mg} where
star distributions in the chemical plane of [M/H] versus [Mg/Fe] are also presented in the bottom panel).
Following J18, a solid line was used to cut the sample into a high-[Mg/Fe] sample and a low-[Mg/Fe] sample, 
corresponding to a thick disk star sample and a thin disk star sample, respectively.

\section{Method and results}

\begin{figure*}
        \centering
        \includegraphics[width=\hsize]{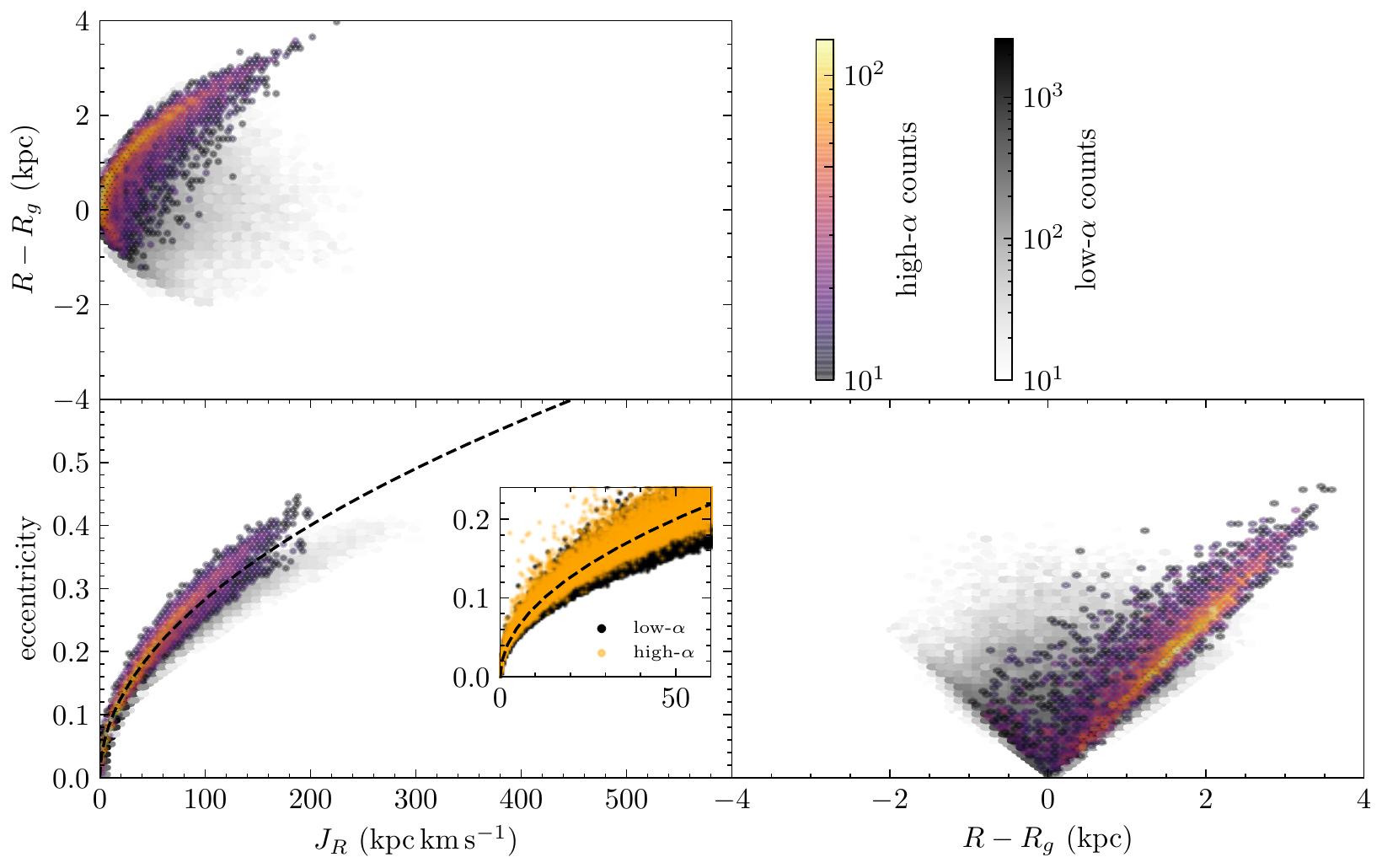}
        \caption{Relations among radial action $J_R$, eccentricity, and $R-R_g$. The black dashed line in the lower-left panel represents the relation between eccentricity and radial action for stars with a guiding radius of 8 kpc and a radial frequency of 39 $\rm km\, s^{-1}\, kpc^{-1}$ derived from epicycle approximation. \label{actions}}
\end{figure*}

As mentioned in the introduction, heating processes increase the random motions of stars and tend to move stars into more eccentric orbits.
We illustrate the relations among the radial actions $J_R$, eccentricities $e,$ and $R-R_g$ of the sample stars in Fig.~\ref{actions}. 
According to the theory of the epicycle approximation, the radial action of a star with a near-circular orbit can be expressed by its eccentricity through $J_R=\frac{E_R}{\kappa}=\frac{1}{2} \kappa e^2 {R_g}^2$, 
where $\kappa$ denotes radial frequency. In the lower-left panel of Fig.~\ref{actions}, a black dashed line
shows the relation between radial action and eccentricity for stars with a guiding radius of 8 kpc and radial frequency of 39 $\rm km\, s^{-1}\, kpc^{-1}$.
Fig.~\ref{actions} shows that kinematically hot stars
tend to have large radial actions and large $|R-R_g|$.

\begin{figure}
        \centering
        \includegraphics[width=\hsize]{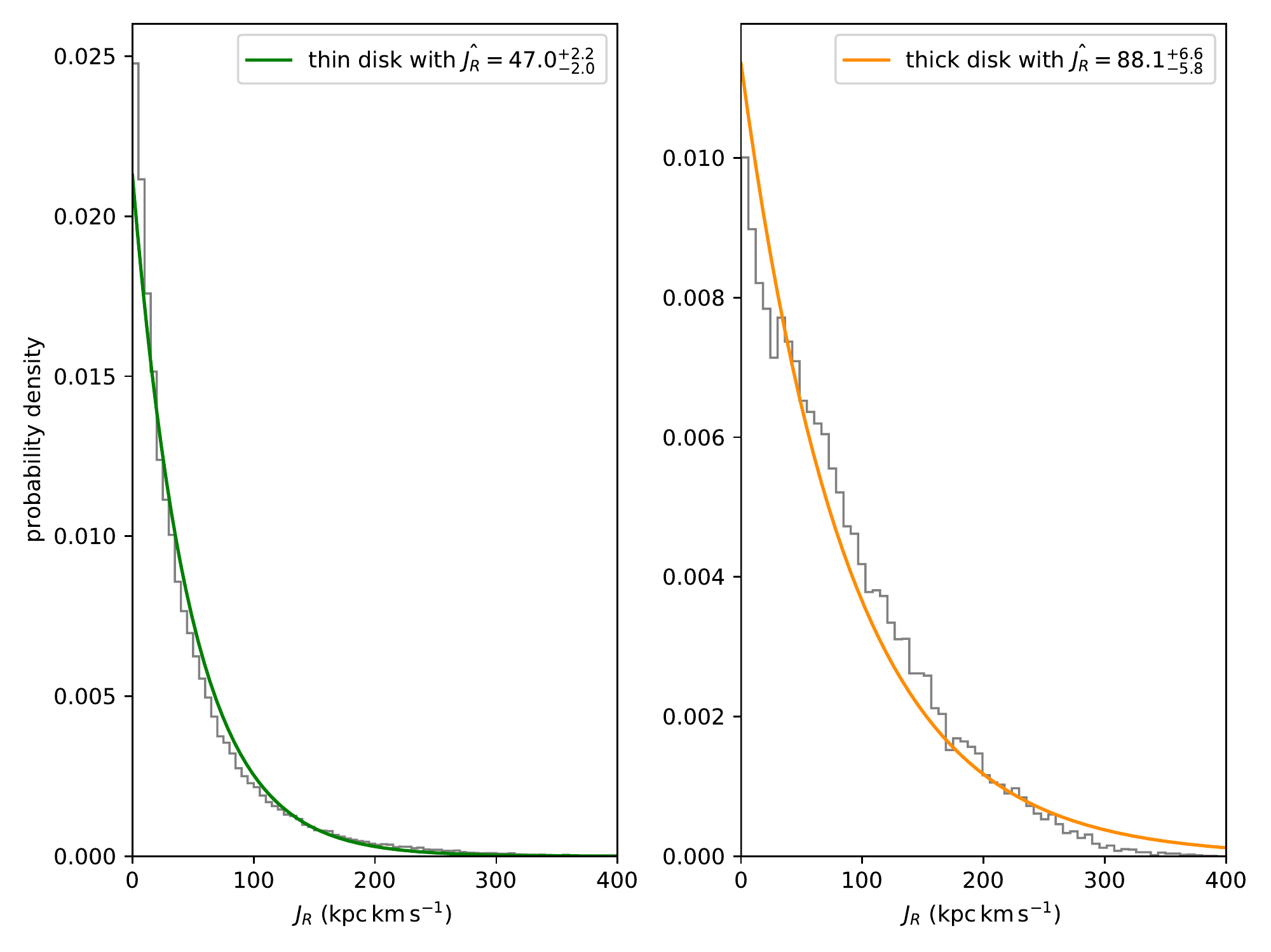}
        \caption{Distributions of radial actions for the thin and thick disks. The solid lines represent the best-fit pseudo-isothermal distribution $J_R\propto exp(-J_R/\hat{J_R})$, where $\hat{J_R}$ is a free parameter to be estimated.  \label{jr-distribution}}
\end{figure}

 Before investigating the relations between radial actions and the vertical distributions for the thin and thick disks, we focused on the
 distributions of radial actions for the thin and thick disks in Fig.~\ref{jr-distribution}. These distributions can be described by pseudo-isothermal distributions \citep{Carlberg1985,Binney2010}, $J_R\propto \exp(-J_R/\hat{J_R})$, where $\hat{J_R}$ is a free parameter standing for the temperature of the disk. We estimated this parameter by using the maximum likelihood technique \citep{Bienayme1987}.
 Specifically, we obtained the ratio of the star number in a given $J_R$ bin and the bin's width and compared them with the simulated data deriving from the above pseudo-isothermal distribution to search 
 for the best-fit $\hat{J_R}$ . The uncertainty was similarly estimated to that of  \citet{Chang2011} and \citet{ChenBQ2017}. 
 
 We calculated the likelihood 500 times using the observed data and the simulations of the best fitting model, applying the individual Poisson noise.
 The resulting likelihood range defined the uncertainty.
 The best-fit $\hat{J_R}$ for each disk is listed in the legends of Fig.~\ref{jr-distribution}, which is 47.0 $\rm kpc\, km\, s^{-1}$ and 88.1 $\rm kpc\, km\, s^{-1}$ for the thin disk and the thick disk, respectively. 
 The pseudo-isothermal distribution of radial action requires that
 the best-fit $\hat{J_R}$ be equal to the mean radial action of the star sample, which in our case is 47.1 $\rm kpc\, km\, s^{-1}$ for the thin disk and 81.9 $\rm kpc\, km\, s^{-1}$ for the thick disk. Regarding the uncertainties of the best-fit $\hat{J_R}$ 
 listed in the legend of Fig.~\ref{jr-distribution}, the best-fit value of $\hat{J_R}$  almost matches the mean radial action of the star sample, indicating that either $\hat{J_R}$ or the mean radial action is a suitable proxy for the temperature of the disk. 
Once a heating process increases the temperature of a star population from a pseudo-isothermal distribution with a low temperature to a high one, the mean radial action (or $\hat{J_R}$) for that star population would also increase.

\subsection{Results from the overall Galactic disk}
\begin{figure*}
\centering
\includegraphics[width=\hsize]{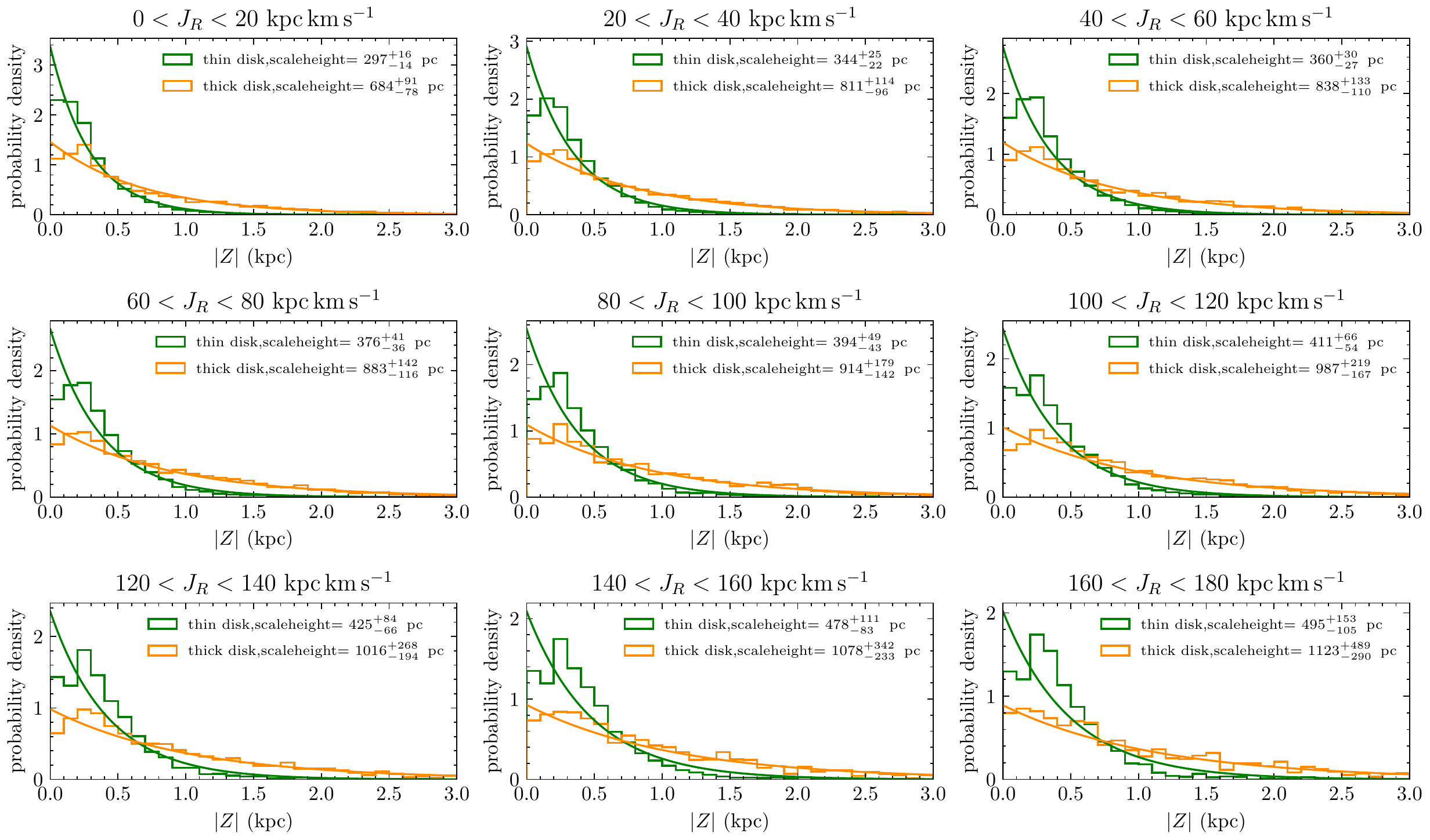}
\caption{Vertical distributions coupled with their best-fit models of the star samples in different radial action ranges. The vertical distributions have been fitted with a simple exponential function with a scale height. The best-fit scale heights are listed in the legends.
In all panels, the solid lines are drawn from the best-fit results.  \label{Z-distribution}}
\end{figure*}

In Fig.~\ref{Z-distribution} we illustrate the vertical distributions in different radial action ranges for both the thin and thick disk star samples.
In order to quantify the relation of radial action with the vertical distribution, every vertical number density distribution was fitted with an exponential function, that is, $\rho(Z) \propto \exp(-|Z|/h)$, where $h$ is the scale height, by using a procedure similar to what we used when fitting the distributions of radial action. We note that, at a large distance, this exponential function is approximate to the isothermal distribution, $\rho(Z) \propto \mathrm{sech}^2(-|Z|/(2h))$.
The resulting best-fit scale heights with their uncertainties are listed in the legend in each of the panels in Fig.~\ref{Z-distribution}. 

\begin{figure}
        \centering
        \includegraphics[width=\hsize]{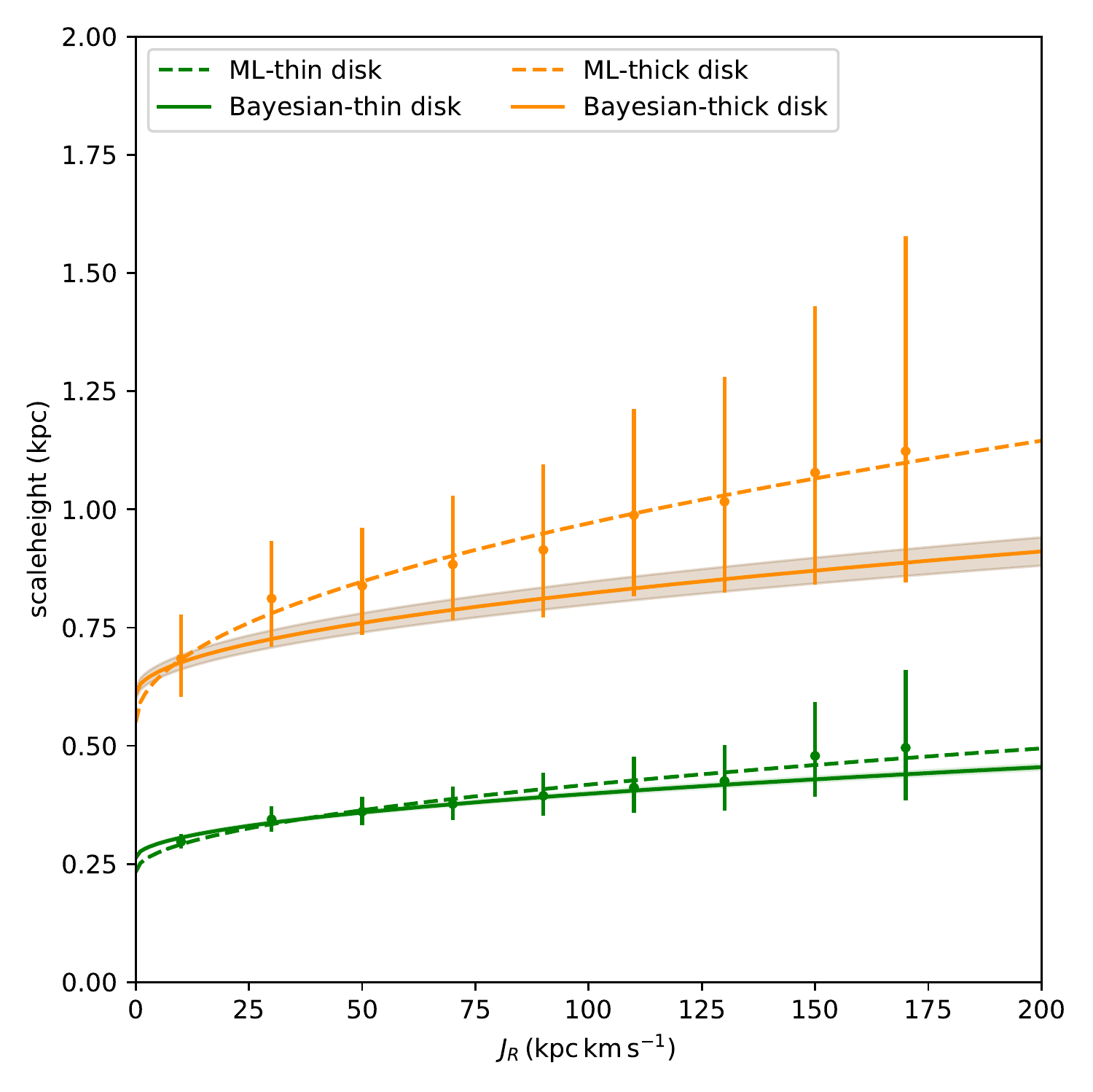}
        \caption{Relations between radial action $J_R$ and scale height $h$ for the thin and thick disks. The relations are fitted with a function of $h=\sqrt{J_R/a}+b$. Both the maximum likelihood (ML) and Bayesian estimations were adopted. The corresponding best-fit relations are indicated by the dashed and solid lines, respectively. The shadowed areas indicate the uncertainties of the Bayesian estimations.}  \label{jr-scale height}
\end{figure}

In Fig.~\ref{jr-scale height} we show the relations between radial action
$J_R$ and scale height $h$ for both the thin and thick disk, which are derived from Fig.~\ref{Z-distribution}. 
We attempt to describe the scale height as a function of radial action with a form of $J_R=a \cdot (h-b)^2$,  where $a$ and $b$ are free parameters. An alternative form of this function is $h=\sqrt{J_R/a}+b$.
The best-fit relations obtained from the maximum likelihood estimation technique are also shown in Fig.~\ref{jr-scale height}. 

Considering that the best-fit scale heights may be less reliable due to the large uncertainties in large radial action bins (especially in the thick disk), which would affect the estimated parameters, we also applied a Bayesian inference.
Compared to the maximum likelihood estimation technique, the advantage of Bayesian inference is that it directly compares the data with the model, avoiding the estimation of the scale height.
By combining the normalized vertical number density of $\rho(Z)=\frac{1}{2h} \exp(-|Z|/h)$ and the relationship of   $h=\sqrt{J_R/a}+b$, we can easily write the logarithm of the probability of finding a star with a given radial action of $J_R$ at distance $Z$, 
\begin{equation}
\ln p(Z|J_R,a,b) = -\ln 2(\sqrt{J_R/a}+b) - \frac{|Z|}{\sqrt{J_R/a}+b}.
\end{equation}
Therefore, the probability of finding a star where the radial action is $J_R$ and the distance is $Z$ can be expressed through the known pseudo-isothermal distribution of radial action($p(J_R) = \exp(-J_R/\hat{J_R})/\hat{J_R}$), that is:
\begin{equation}
p(Z,J_R|a,b)=p(Z|J_R,a,b) \cdot p(J_R) .\nonumber
\end{equation}
By taking the logarithm of the above equation, we obtained the likelihood function: 
\begin{equation}
\ln p(Z,J_R|a,b) = -\ln 2(\sqrt{J_R/a}+b) - \frac{|Z|}{\sqrt{J_R/a}+b} - \ln \hat{J_R} - \frac{J_R}{\hat{J_R}}.
\end{equation}
Then assuming a uniform prior that requires $a>0$ and $b>0$, we formulated the logarithm of posterior (up to a constant) to be:
\begin{equation}
\ln p(a,b|\{Z,J_R\}) = \sum_{i=1} -\ln 2(\sqrt{J_{R,i}/a}+b) - \frac{|Z_i|}{\sqrt{J_{R,i}/a}+b} - \ln \hat{J_R} - \frac{J_{R,i}}{\hat{J_R}}.
\end{equation}
The summation is over all stars, and the value of $\hat{J_R}$ was chosen to be the mean of radial action of the star sample, instead of the best-fit values shown in Fig.~\ref{jr-distribution} for the thin and thick disks. We find that this choice does not affect the estimated best-fit parameters.

\begin{figure*}
        \centering
        \includegraphics[width=0.49\hsize]{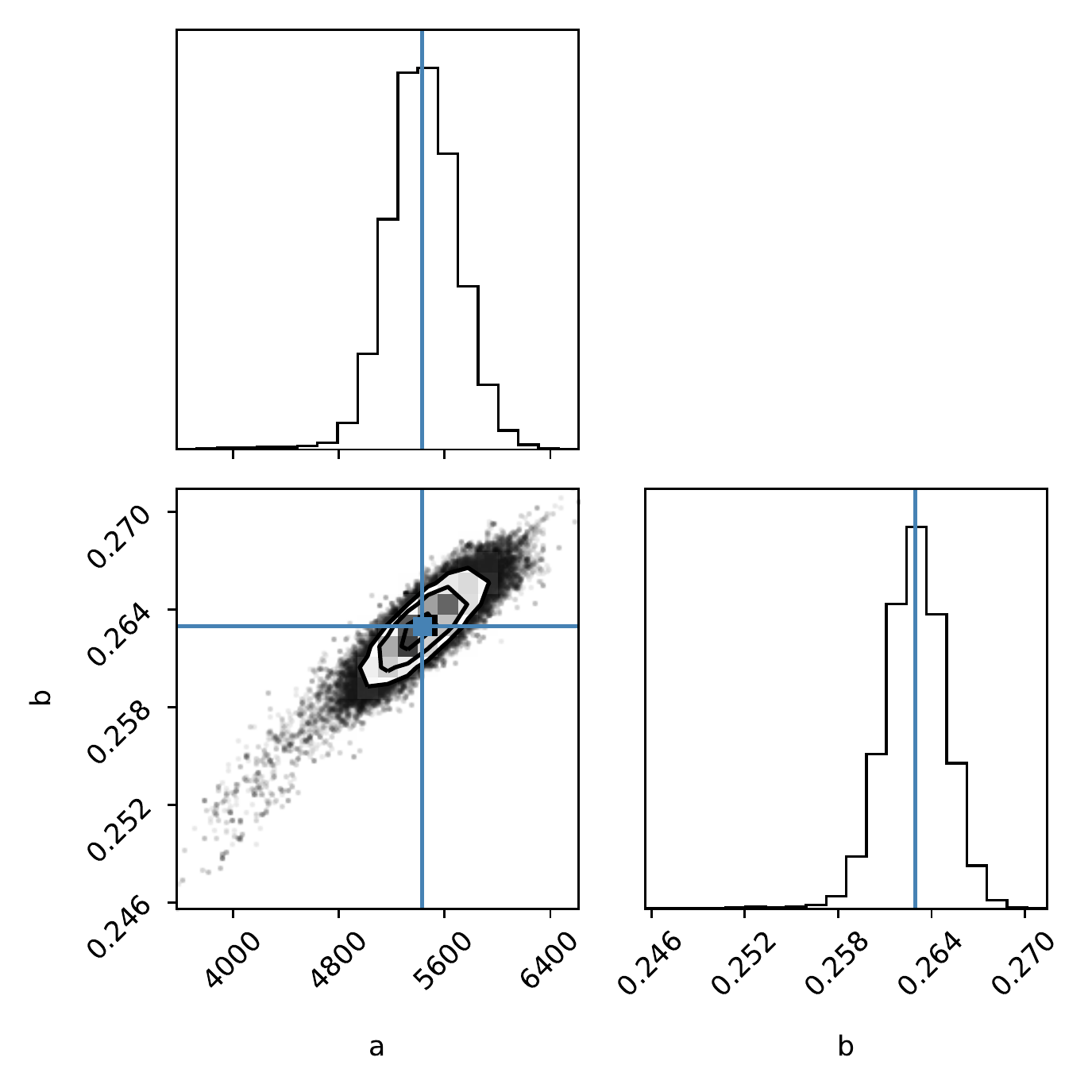}
        \includegraphics[width=0.49\hsize]{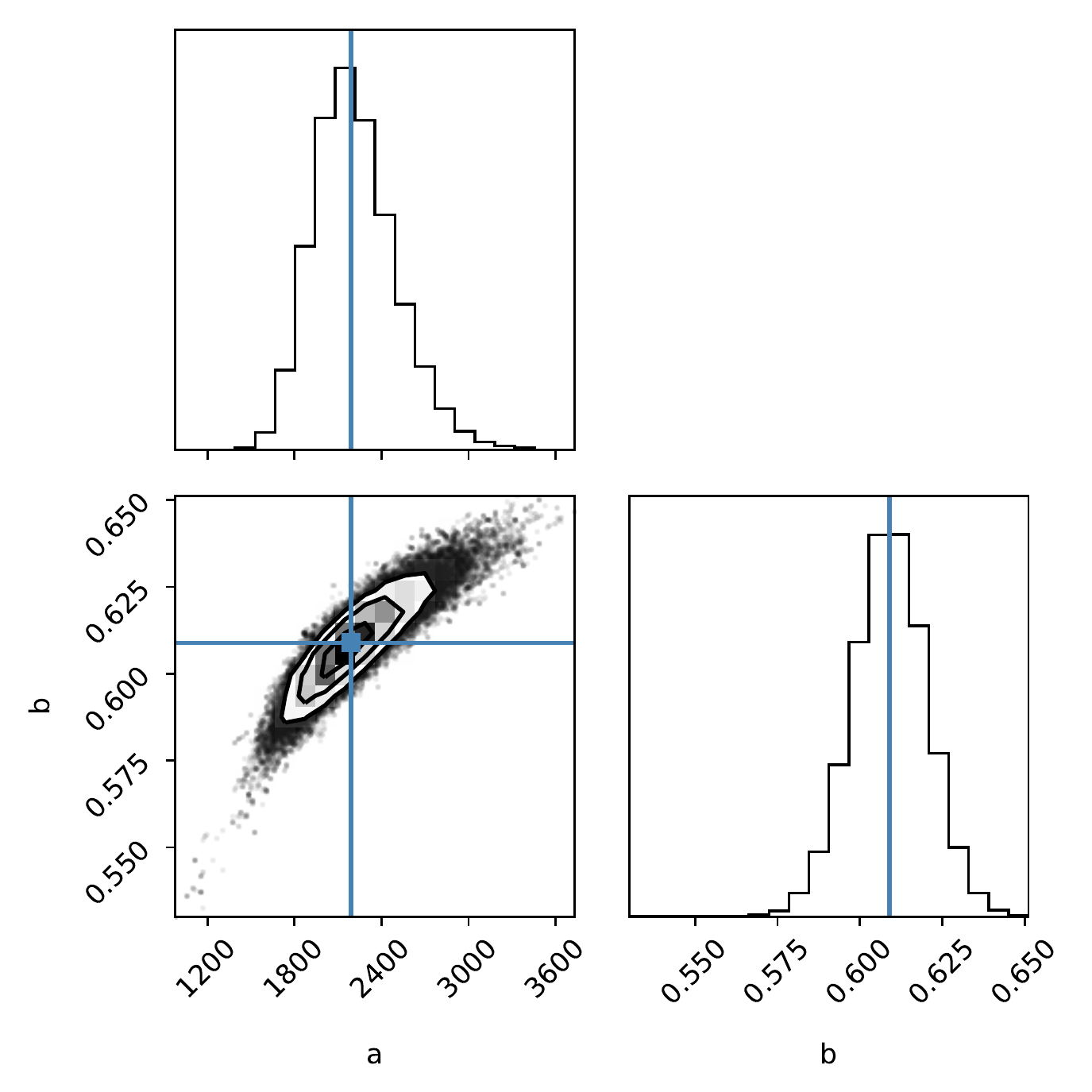}
        \caption{Posterior of the model parameters for the overall thin (left panel) and thick (right panel) disks. The medians of the parameters are shown by blue lines.\label{coner-plot}}
\end{figure*}

We sampled the posterior via the EMCEE package with 1,000 steps and 100 walkers. The initial start points of the chain are just the best-fit parameters that were obtained from the maximum likelihood estimation. Fig.~\ref{coner-plot} illustrates the sampling of the model parameter posterior after discarding the initial 100 steps, and we used the 16th, 50th, and 84th percentiles to quote the uncertainties of the parameters. The best-fit relations from Bayesian inference are
\begin{equation}
\begin{aligned}
h =& \sqrt{J_R/5431.89}+0.263, \rm{thin\, disk} \\
h =& \sqrt{J_R/2190.09}+0.609, \rm{thick\, disk.}
\end{aligned}
\end{equation}
Predictions from these best-fit results are shown in Fig.~\ref{jr-scale height} as are the uncertainties in predictions. 

Figure~\ref{jr-scale height} clearly shows that the maximum likelihood (or the data points) and Bayesian estimations give a consistent relationship for the thin disk, while they present a discrepancy for the thick disk. This discrepancy might arise from the estimated scale heights of the thick disk in large radial action ranges that are less reliable, reflected by their large uncertainties in Fig.~\ref{Z-distribution}. However, when taking the uncertainties of the Bayesian prediction and  the best-fit scale heights into account, the resulting relation from Bayesian inference is basically in line with the data points. Considering that the Bayesian inference directly compares the model with data, 
we favor the results derived from this method.
 
\begin{figure}
        \centering
        \includegraphics[width=\hsize]{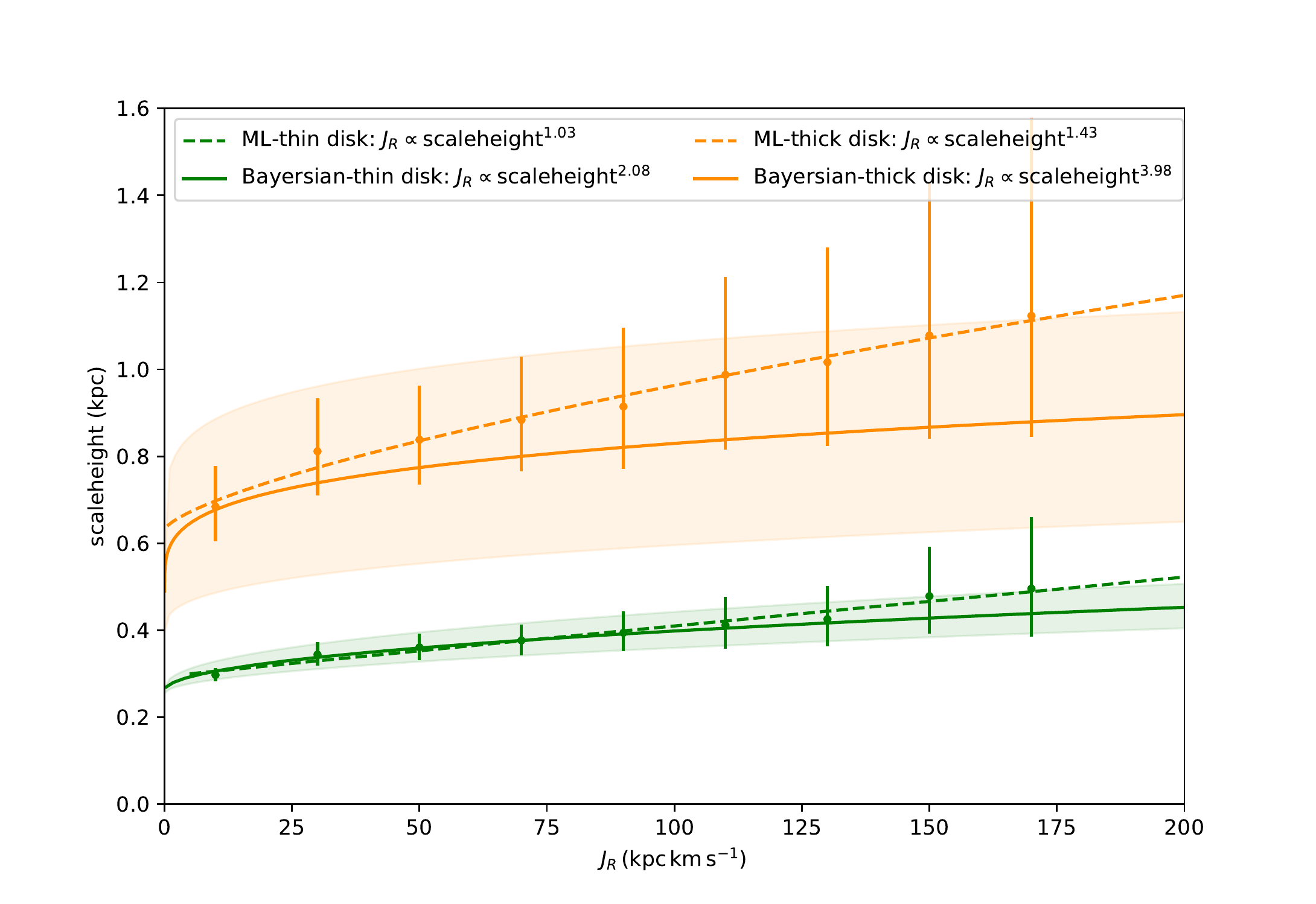}
        \caption{Testing a three-parameter function ($J_R=a \cdot (h-b)^\alpha$ or $h=\sqrt[\alpha]{J_R/a}+b$) to describe the relationships between radial action $J_R$ and scale height $h$ for the thin and thick disks. As  done in Fig.~\ref{jr-scale height},  the maximum likelihood (ML) and Bayesian estimations were both applied. The best-fit value for $\alpha$ can be found in the legend. The results from Bayesian inference are more favored, as this method directly compares the model with data.  \label{three-parameters}}
\end{figure}

We also explored a three-parameter function of $J_R=a \cdot (h-b)^\alpha$ to describe the relationship between  radial action and scale height. In the case where $\alpha=2$, this three-parameter function is the same model that we used in Fig.~\ref{jr-scale height}.
The best-fit results of this three-parameter model are illustrated in Fig.~\ref{three-parameters}. The best-fit $\alpha$ estimated from Bayesian inference is $2.08^{+0.14}_{-0.16}$ for the thin disk, which is highly consistent with the function that was used in Fig.~\ref{jr-scale height}. For the thick disk, the best-fit $\alpha$ is $3.98^{+1.47}_{-1.08}$, which might indicate that the function used in Fig.~\ref{jr-scale height} is inappropriate for the thick disk. The uncertainties of $\alpha$ and of the prediction from  Bayesian inference are too large for the thick disk,
reducing the reliability of this three-parameter function.
Thus, we prefer to still use the two-parameter function ($J_R=a \cdot (h-b)^2$) to describe the relationship between radial action and scale height for the thick disk in this work,
but we note that this two-parameter function for the thick disk should be  used carefully in other works.

To better understand the relationships between the radial action and scale height we obtained, we started from the isothermal limit and epicycle approximation.
In the isothermal limit \citep[e.g., Problems 4.21 in][]{binney2008-book} and epicycle approximation, the radial action is $J_R=E_R/\kappa = \sigma_R^2/\kappa$ and $\sigma_z^2=8\pi G\rho_0 h^2$. To express radial action $J_R$ in terms of scale height required $\sigma_R$ to be a known function of $\sigma_z$. In our case, $J_R \propto (h-b)^2$ requires $\sigma_R^2 \propto (\sigma_z-b\sqrt{8\pi G\rho_0})^2$. 
Therefore, the existence of the relationship between radial action and scale height reveals a fixed relationship between $\sigma_R$ and $\sigma_z$.
Without the knowledge of the exact relation between $\sigma_R$ and $\sigma_z$ from other data or theory to justify the relationship we measured, we elected to use $J_R=a \cdot (h-b)^2$ to describe the relationship between radial action and scale height for both of the disks.

\subsection{Results from the inner and outer disk}
Previous works have found that some of the properties of the inner disk are different from the outer disk \citep[e.g.,][]{Haywood2013,Haywood2016,Mackereth2019},  which seems to
indicate different evolutionary histories. By using the guiding radius, we separated the  Galactic disks into an inner disk set and an outer disk set. We defined the stars with a guiding radius smaller than 8 kpc as being an inner disk, and those with a guiding radius larger than 8 kpc 
as being an outer disk. The reason for using the guiding radius instead of the current galactocentric radius was that the guiding radius is more suited to representing the position of a star's orbit.

\begin{figure*}
        \centering
        \includegraphics[width=\hsize]{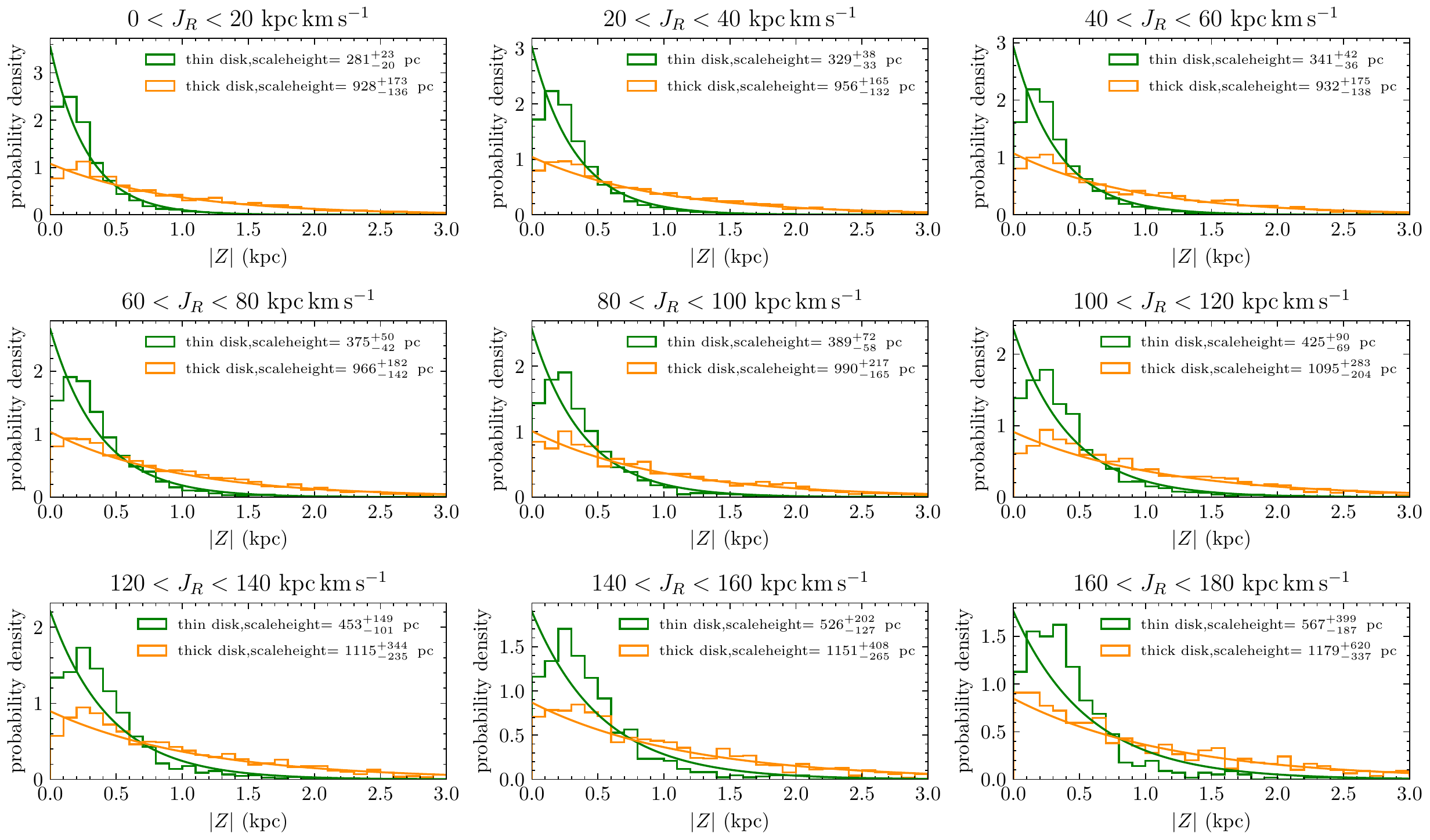}
        \caption{Vertical distributions coupled with their best-fit models of the star samples in different radial action ranges in the inner disk($R_g < 8$ kpc), as done in Fig.~\ref{Z-distribution}. \label{inner}}
\end{figure*}

\begin{figure*}
        \centering
        \includegraphics[width=\hsize]{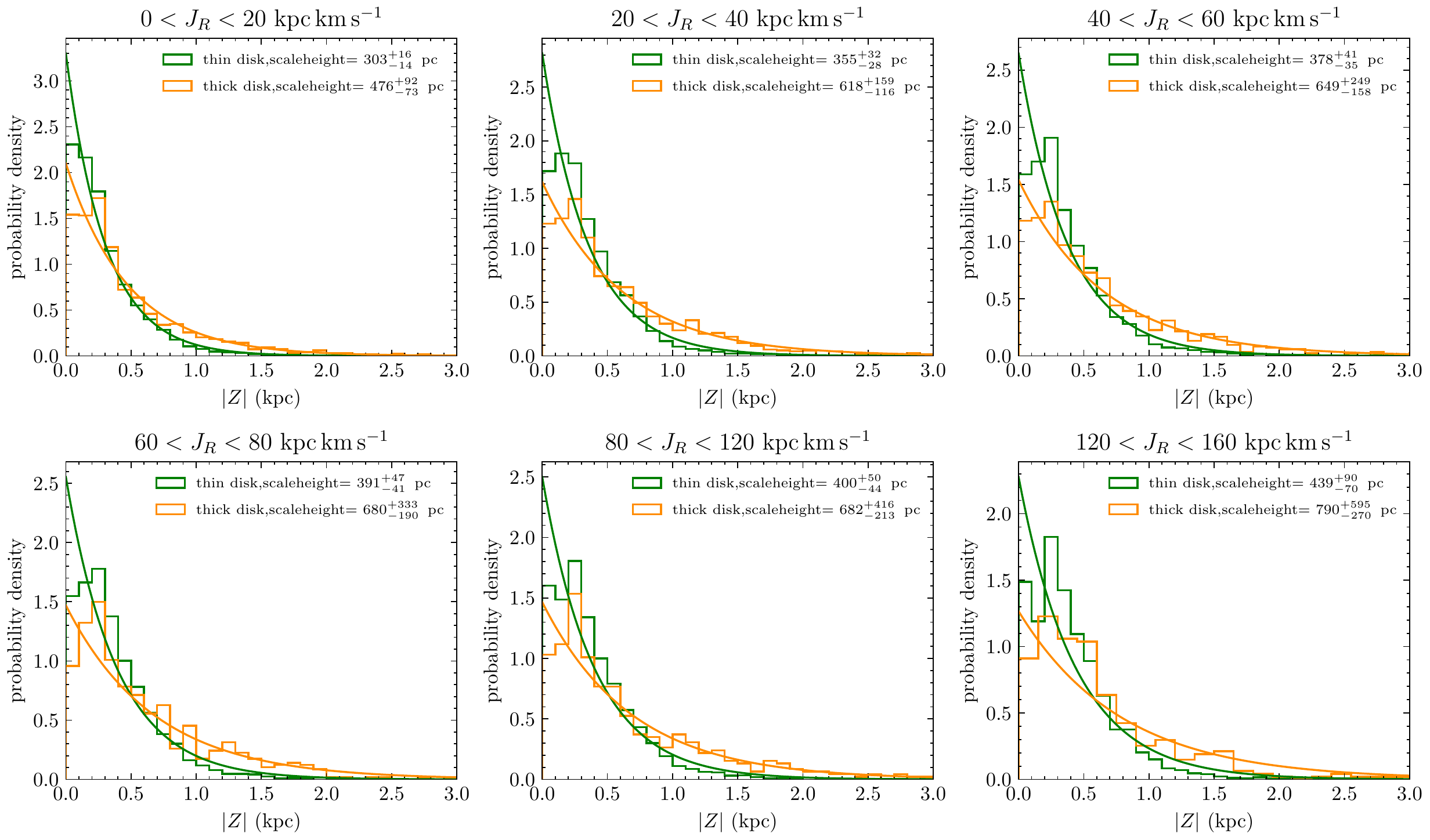}
        \caption{Vertical distributions coupled with their best-fit models of the star samples in different radial action ranges in the outer disk($R_g > 8$ kpc), as done in Fig.~\ref{Z-distribution}. It should be noted that we increase the size of radial action ranges in the last two lower panels, as the number of stars is relatively small in the outer disk, and the thick disk gives a relatively poor fit in the lower-right panel due to the insufficient number of stars in this radial action range.   \label{outer}}
\end{figure*}

 We obtained the vertical distributions of the defined inner and outer disks and their corresponding best-fit distributions as we had done in Fig.~\ref{Z-distribution}, which are illustrated in Fig.~\ref{inner} and \ref{outer}, respectively.
We point out that we increased the size of the radial action ranges in the last two lower panels of Fig.~\ref{outer}, as the number of stars in the outer disk is relatively small. 

\begin{figure*}
        \centering
        \includegraphics[width=0.49\hsize]{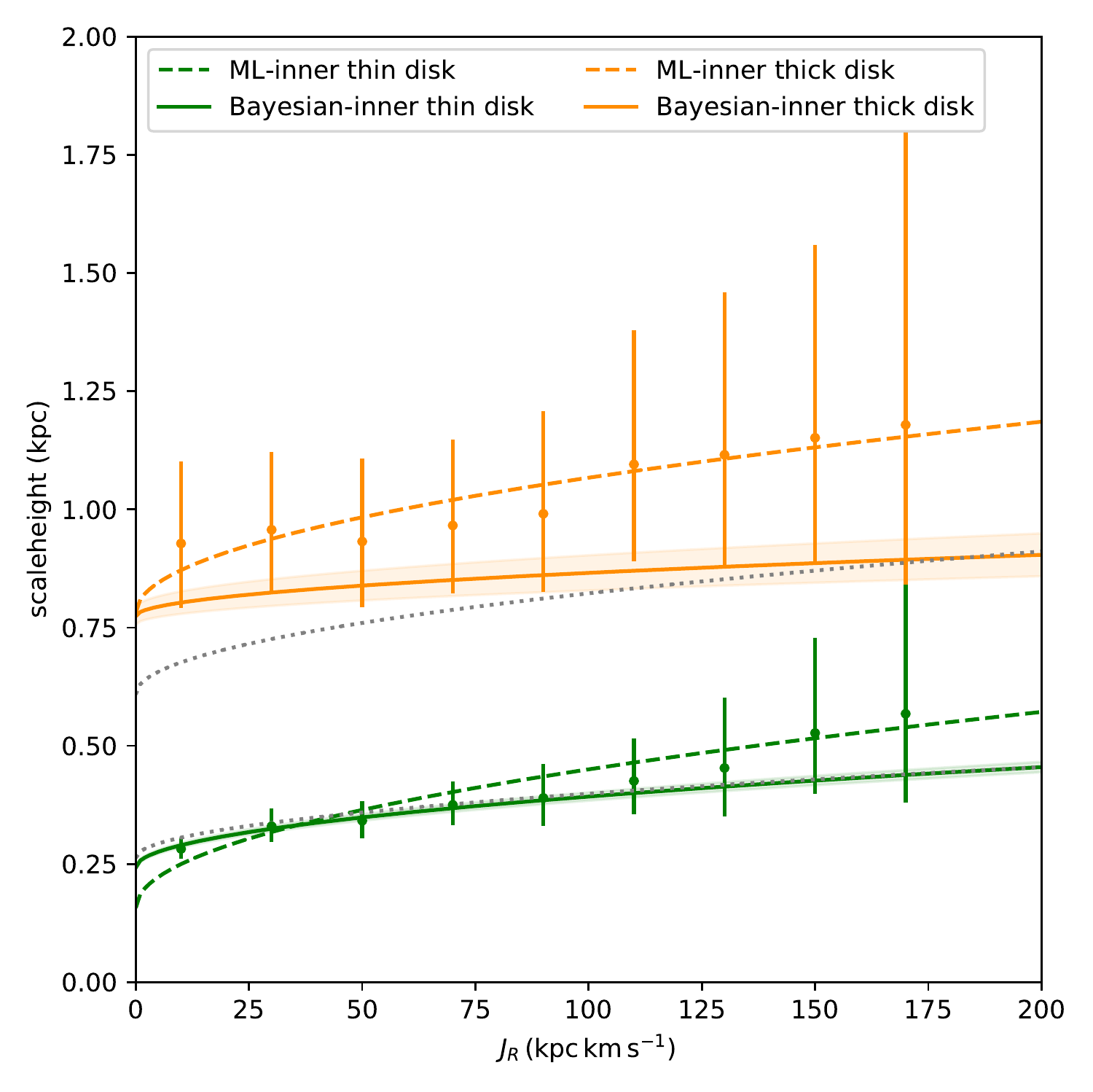}
        \includegraphics[width=0.49\hsize]{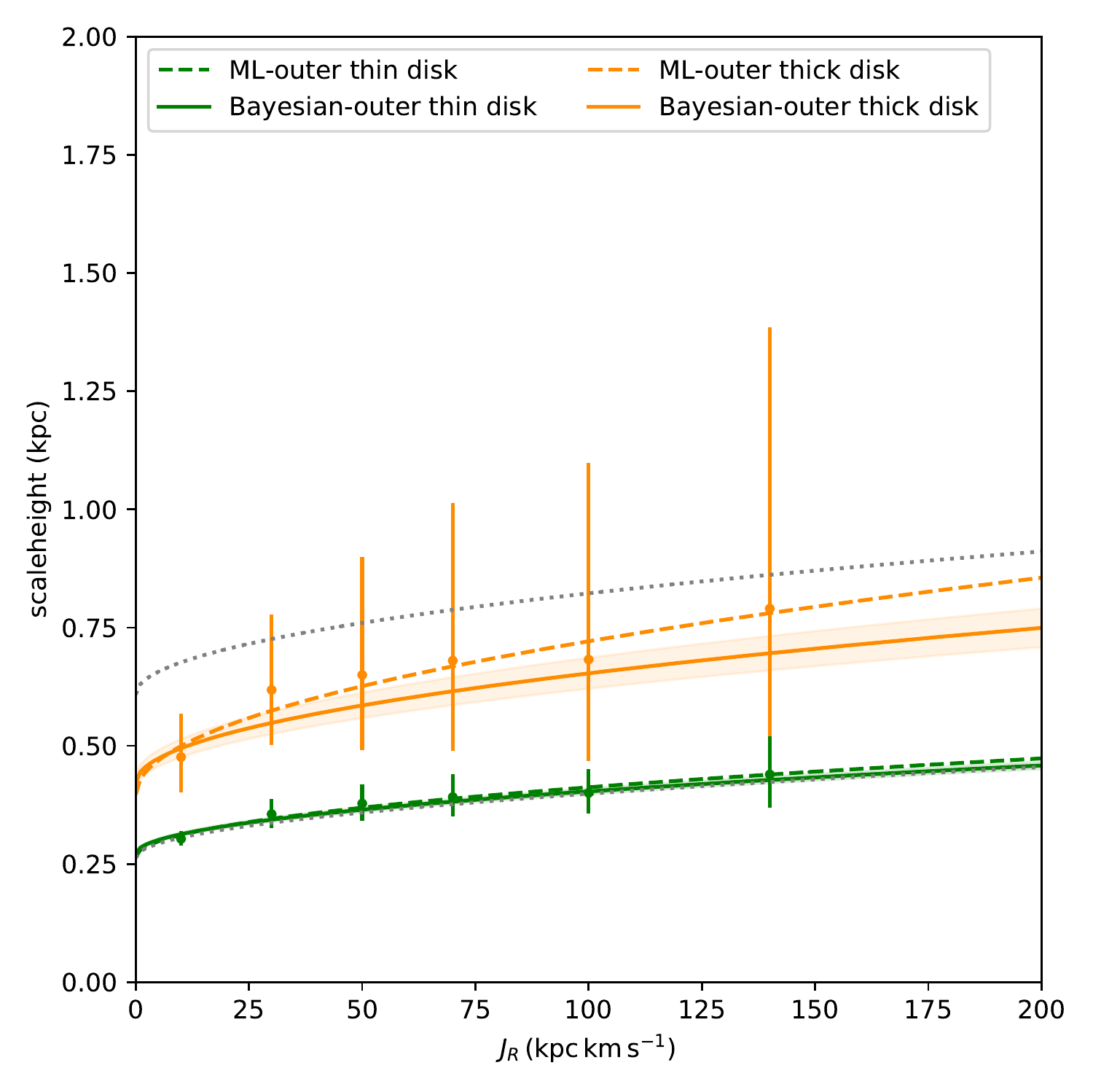}
        \caption{Relationships between radial action $J_R$ and scale height $h$ in the inner(left panel) and outer(right panel) disk. The relationships are fitted with a function of $h=\sqrt{J_R/a}+b$, as done in Fig.~\ref{jr-scale height}. The best-fit relationships estimated from Bayesian inference for the overall thin and thick disk star samples in Fig.~\ref{jr-scale height} are also plotted for comparison as  dotted gray lines.  \label{inner-outer}}
\end{figure*}

The relationships between radial actions and scale heights for the inner and outer disks are shown in Fig.~\ref{inner-outer}. As done in Fig.~\ref{jr-scale height}, both maximum likelihood and Bayesian estimations were applied, but the latter is preferred, as we explained in the previous subsection. The best-fit relations derived from Bayesian inference for the overall thin and thick disk star samples in Fig.~\ref{jr-scale height} are also plotted for comparison. The inner and outer thin disks show nearly the same relations as that of the overall thin disk. However, the inner and outer thick disks present different relationships from each other. The inner thick disk has a relatively flat relation, and the scale heights are larger than those of the overall thick disk, except at the large radial action end. In contrast, the outer thick disk shows a similar trend with that of the overall thick disk, but the scale height is smaller at all radial actions.

\section{Discussion and conclusions}
In this work, we have analyzed the scale height $h$ as a function of radial action $J_R$
for the overall disk in Fig.~\ref{jr-scale height} and for the inner and outer disk in Fig.~\ref{inner-outer} via two different approaches: maximum likelihood and Bayesian estimations. In order to directly compare this function with data, we estimated the scale heights in radial action ranges and then described the relationship between the scale height and radial action with a function of  $h=\sqrt{J_R/a}+b$ using the maximum likelihood estimation technique. As the estimated scale heights for the thick disk at large radial action ranges are too large and may affect the best-fit relationship between radial action and scale height, we also applied a Bayesian estimation to directly compare the model with data, avoiding to estimate the scale height in different radial action ranges. It should be noted that we do not correct any biases in this work.
The reasons are: 1) we only used chemical information to divide the thin and thick disks;
2) neither the APOGEE nor Gaia data are kinematically biased \citep{Mackereth2019}; and
3) most mono-age, mono-[Fe/H] populations are close to isothermal  \citep[e.g.,][]{Mackereth2019}.
Therefore, the disk stars  are mostly phase-mixed in the vertical direction.
The first two reasons demonstrate that the only bias arises from APOGEE spatial selection. However, the third reason assumes the bias does not strongly affect our main results, as we are only interested in the vertical distributions of the disk stars.

We find that a function of $h=\sqrt{J_R/a}+b$ is able to describe the thin disk well, but it should be used with care for the thick disk.
The best-fit function given by Bayesian inference is 
$h =\sqrt{J_R/5431.89}+0.263$ for the thin disk, whereas 
$h = \sqrt{J_R/2190.09}+0.609$ for the thick disk.
Moreover, we find that the thin disk has nearly the same relationship between the inner and outer disk, as seen from Fig.~\ref{inner-outer}, while the thick disk presents different relationships: the inner thick disk shows a nearly flattened trend, and the outer thick disk has a gradually increasing trend.

As heating agents can broaden the distribution of radial action and increase the scale height of disk stars, the measured relationships encode the heating history and the origin information of the star samples.
In the isothermal limit and epicycle approximation, we show that a function of $h=\sqrt{J_R/a}+b$ reveals a fixed relationship between the radial and vertical velocity dispersion, that is, $\sigma_R^2 \propto (\sigma_z-b\sqrt{8\pi G\rho_0})^2$. This relationship is the overall consequence from all of the heating agents.
Without the aid of a simulation, it is hard to discuss the implications of these relationships on the formation and heating history of the disks.  However, it is reasonable to suspect that giant molecular clouds may be a dominant or non-negligible heating agent to account for in these relationships, as they heat both radially and vertically and hence have a chance to form a fixed relationship between $\sigma_R^2$ and $\sigma_z^2$.
We would like to further explore this content in the future.
Most importantly, this work provides an alternative way to challenge models and thus study the heating history of the Galactic disks by investigating the relationship between radial action and scale height.

\begin{acknowledgements}
It is a pleasure to thank the referee for valuable comments and helpful suggestions which improve the manuscript much.
This work is supported by the National Natural Science Foundation of China (Grant Nos. 11988101, 11890694, 11973042, 11973052), the China Manned Space Project with No. CMS-CSST-2021-B05, National Key R\&D Program of China (Grant No. 2019YFA0405502) and the 2-m Chinese Space Survey Telescope project.

Funding for the Sloan Digital Sky Survey IV has been provided by the Alfred P. Sloan Foundation, the U.S. Department of Energy Office of Science, and the Participating Institutions. SDSS-IV acknowledges
support and resources from the Center for High-Performance Computing at
the University of Utah. The SDSS web site is www.sdss.org.

SDSS-IV is managed by the Astrophysical Research Consortium for the
Participating Institutions of the SDSS Collaboration including the
Brazilian Participation Group, the Carnegie Institution for Science,
Carnegie Mellon University, the Chilean Participation Group, the French Participation Group, Harvard-Smithsonian Center for Astrophysics,
Instituto de Astrof\'isica de Canarias, The Johns Hopkins University,
Kavli Institute for the Physics and Mathematics of the Universe (IPMU) /
University of Tokyo, Lawrence Berkeley National Laboratory,
Leibniz Institut f\"ur Astrophysik Potsdam (AIP),
Max-Planck-Institut f\"ur Astronomie (MPIA Heidelberg),
Max-Planck-Institut f\"ur Astrophysik (MPA Garching),
Max-Planck-Institut f\"ur Extraterrestrische Physik (MPE),
National Astronomical Observatories of China, New Mexico State University,
New York University, University of Notre Dame,
Observat\'ario Nacional / MCTI, The Ohio State University,
Pennsylvania State University, Shanghai Astronomical Observatory,
United Kingdom Participation Group,
Universidad Nacional Aut\'onoma de M\'exico, University of Arizona,
University of Colorado Boulder, University of Oxford, University of Portsmouth,
University of Utah, University of Virginia, University of Washington, University of Wisconsin,
Vanderbilt University, and Yale University.

This work has made use of data from the European Space Agency (ESA) mission
{\it Gaia} (\url{https://www.cosmos.esa.int/gaia}), processed by the {\it Gaia}
Data Processing and Analysis Consortium (DPAC,
\url{https://www.cosmos.esa.int/web/gaia/dpac/consortium}). Funding for the DPAC
has been provided by national institutions, in particular the institutions
participating in the {\it Gaia} Multilateral Agreement.

\end{acknowledgements}

%
%
\bibliographystyle{aa} 
\bibliography{ref} 

\end{document}